\documentclass[openacc]{rstransa}

\def \aj {AJ}
\def \mnras {MNRAS}

\def \apj {ApJ}

\def \nat {Nature}
\def \araa {ARAA}

\def \aapr {A\&A Rev.}
\def\lesssim{\mathrel{\hbox{\rlap{\hbox{\lower4pt\hbox{$\sim$}}}\hbox{$<$}}}}

\begin{document}

\title{Bridging the Gap: From Massive Stars to Supernovae}

\author{
Justyn R. Maund$^{1,2}$, Paul A. Crowther$^{1}$, Hans-Thomas Janka$^{3}$ and Norbert Langer$^{4}$}

\address{$^{1}$Department of Physics and Astronomy, University of Sheffield, Hicks Building, Hounsfield Road, Sheffield, S3 7RH, U.K.\\
$^{2}$Royal Society Research Fellow\\
$^{2}$Max-Planck-Institut f\"{u}r Astrophysik, Karl-Schwarzschild-Str. 1, D-85748 Garching, Germany\\
$^{3}$Argelander-Institut f\"{u}r Astronomie der Universit\"{a}t Bonn, Auf dem H\"{u}gel 71, 53121 Bonn, Germany}

\subject{Astrophysics}


\corres{Justyn R. Maund\\
\email{j.maund@sheffield.ac.uk}}

\begin{abstract}
Almost since the beginning, massive stars and their resultant supernovae have played a crucial role in the Universe.  These objects produce tremendous amounts of energy and new, heavy elements that enrich galaxies, encourage new stars to form and sculpt the shapes of galaxies we see today.  The end of millions of years of massive star evolution and the beginning of hundreds or thousands of years of supernova evolution are separated by a matter of a few seconds, in which some of the most extreme physics found in the Universe causes the explosive and terminal disruption of the star.   Key questions remain unanswered in both the studies of how massive stars evolve and the behaviour of supernovae, and it appears the solutions may not lie on just one side of the explosion or the other or in just the domain of the stellar evolution or the supernova astrophysics communities.  The need to view massive star evolution and supernovae as continuous phases in a single narrative motivated the Theo Murphy international scientific meeting ``Bridging the gap: from massive stars to supernovae" at Chicheley Hall in June 2016, with the specific purpose to simultaneously address the scientific connections between theoretical and observational studies of massive stars and their supernovae, through engaging astronomers from both communities.
\end{abstract}
\maketitle

\begin{fmtext}

\end{fmtext}
\section{Introduction}
Core-collapse supernovae are the expected fate for nearly all massive stars with initial masses $>8M_{\odot}$.  Over their life times, of up to a few tens of millions of years, the energy produced from the fusion of lighter elements to make heavier elements holds these heavy weight stars up against their own gravity \cite{2002RvMP...74.1015W}.  The story of the lives of these stars will be influenced by their interactions with binary companions and how they lose mass through strong but steady winds or violent, but short lived eruptive episodes \cite{2014ARA&A..52..487S}.  The culmination of massive star evolution, when their core supply of fuel is finally exhausted with the production of an Fe core with $M > 1.3M_{\odot}$, is the inevitable collapse of the core to form a neutron star or black hole \cite{2002RvMP...74.1015W}.  The gravitational energy released from this collapse ($\mathrm{\sim 10^{46}J}$) is, through one of a number of  hotly debated physical processes, coupled to the outer layers of the star, exploding it outwards at speeds of $\sim 10^{4}\,\mathrm{km\,s^{-1}}$ \cite{fili97} and with brightnesses $M_{V} \sim -17  \,\mathrm{mags}$ \cite{1990AJ....100..530M}.  For the vast majority of core-collapse supernovae their brightness will be dictated by the decay of heavy, radioactive elements (in particular $\mathrm{^{56}Ni}$; \cite{1969ApJ...157..623C}) forged in the high densities and temperatures of the explosion.   Eventually, after a period of $\sim 10^{1}$ years, the expanding ejecta will  transition into a remnant, which ploughs into and eventually merges with the interstellar medium.\\

Massive stars and their supernovae play an important role in the Universe and in the field of astronomy itself: 
\begin{enumerate}
\item{The strong winds and ionising radiation of massive stars disperse products from stellar nucleosynthesis and ionise their host galaxies \cite{2003ARA&A..41...15M}.}
\item{The cool winds of the red supergiant phase of some massive stars and the cooling remnants of supernovae are the environments in which significant quantities of dust are formed in the Universe \cite{2011A&ARv..19...43G}.}
\item{Core-collapse supernova explosions disperse the nucleosynthetic products built up over the lives of massive stars and, in the high densities and temperatures of the explosion, even heavier elements beyond $\mathrm{^{56}Fe}$ are created.  These heavy elements are important ingredients for new stars to form, for planets and, ultimately, for life \cite{1960ApJ...132..565H}.}  
\item{The huge radiative and kinetic energies of these explosions stir up  their host galaxies, sculpting their shapes, inducing new stars to form and act as agents against the gravitational influence of dark matter \cite{2012MNRAS.421.3464P}.}  
\item{Massive stars and their supernovae are associated with the formation of exotic compact objects, such as neutron stars and black holes. \cite{1934PhRv...46...76B}.}
\item{Some massive star supernovae also have the utility of being useful distance indicators, whose brightness allows them to serve as cosmic yard sticks for measuring the shape of the Universe \cite{1974ApJ...193...27K}. }  
\item{The violent explosion of many solar masses of material at very high velocities also means that core-collapse supernovae are likely to be a key source of gravitational waves\cite{2009CQGra..26f3001O} and, with the detection of neutrinos \cite{2012ARNPS..62...81S}, they will be significant events for ``multi-messenger astronomy" as the field moves beyond merely the measurement of just electromagnetic radiation.}
\end{enumerate}

The nature of massive stars and their supernovae is one of the key science drivers behind the next generation of astronomical facilities, including the Extremely Large Telescopes (ELTs; such as the European ELT \footnote{https://www.eso.org/sci/facilities/eelt/science/doc/eelt\_sciencecase.pdf}, the Giant Magellan Telescope and the Thirty Meter Telescope), the James Webb Space Telescope\footnote{https://www.nasa.gov/pdf/715962main\_jwst\_science\_pub-v1-2.pdf} and, most importantly, the Large Synoptic Survey Telescope\footnote{https://www.lsst.org/scientists/scibook}\cite{lsstsci}.\\  
\section{Mind the Gap}
Despite the importance of these two classes of objects, there are significant questions about both massive stars and supernovae and how the studies of the two classes of objects can be  connected.  Beyond this basic picture outlined above, mapping the complex characteristics of the types of massive stars observed in the Milky Way  and nearby galaxies with the properties of supernovae observed across the Universe, and vice versa, remains a major challenge.  
Even when the individual stars about to explode are observed, by luck, in pre-explosion observations of the supernova site, their stellar properties are often in conflict with those expected from models of massive star evolution and supernovae \cite{2009ARA&A..47...63S}.  Certainly the hope of a one-to-one mapping of a star with a given initial mass to a single supernova type is too simplistic.  This complex and, currently, incomplete mapping is in part due to outstanding problems in both the fields of massive star evolution (e.g. rotation, convection, binarity, mass loss and the role of metallicity etc.; \cite{2002ESASP.485...57W, 2012ARA&A..50..107L}) and supernovae (e.g. the 3D shapes, the source of energy, the amount of unexcited, invisible material etc.; \cite{2008ARA&A..46..433W}), as well as the ways in which the two types of objects can be observed.  The difference in the physical processes associated with the evolution of massive stars and supernovae, from how they are powered to their structure and the different timescales over which they evolve, also means there is a gap in how these two types of objects are modelled and the types of observations that can be made.  While entire populations of massive stars can be observed and spatially resolved in the Milky Way and nearby galaxies, these are effectively single, surface snapshots of a point in their relatively long  life spans.  Conversely, each supernova occurs only once\footnote{for bona fide terminal explosions, since some bright transients have been established to be supernova impostors; i.e. non-catastrophic explosions.} and observers need to react as soon as possible to new discoveries, before the supernovae fade and are no longer observable.

The major gap between the studies of massive stars and supernovae, however, is the explosion mechanism.  The mechanism is a discontinuity between massive star evolution and the birth of  supernovae.  The extreme physics associated with the explosion mechanism, including high energies ($10^{44} - 10^{46}\,\mathrm{J}$), high temperatures, high magnetic field strengths ($\sim 10^{11}\,\mathrm{T}$; \cite{2012ARNPS..62..407J}), high densities ($10^{17}\,\mathrm{kg\,m^{-3}}$; \cite{1979ARA&A..17..415B}) and high velocities, over extremely short timescales $\lesssim 5\,\mathrm{s}$, separate the quasi-equilibrium of massive star evolution of millions of years and the long term expansion of the SN observable over $10\,\mathrm{s} - 100\,\mathrm{s}$ of days and, for nearby cases, into the remnant phase for 1000s of years.  As a further complication, models predict that not all massive stars are able to explode as supernovae and there is a hunt for stars possibly disappearing through failed supernovae \cite{2014ApJ...783...10S, 2008ApJ...684.1336K}.  The nature of the explosion mechanism is, currently, the domain of theoretical models, for which effects due to neutrinos, strongly contorted magnetic fields and hydrodynamical instabilities might all be significant components \cite{2012ARNPS..62..407J}. The specific physics of the explosion are crucial for understanding the origins of the properties of supernovae, in particular their nucleosynthetic signature for establishing the history of the enrichment of the Universe with heavy elements \cite{2016Natur.531..610J}. The growth in observational monitoring campaigns of supernovae in the last decade has not yielded a commensurate increase in diagnostic capability for probing the nature of the explosion mechanism.\\
\section{Bridging the Gap}
Over two days (1 - 2 June 2016), a group of 72 astronomers, covering a range of subjects concerned with massive star evolution and supernovae, from both observational and theoretical perspectives, converged on Chicheley Hall for the Theo Murphy international scientific meeting ``Bridging the gap: from massive stars to supernovae".  Unlike previous meetings, which have either been primarily devoted to massive stars or to supernovae (with a small portion given over to the other subject), for this meeting it was purposefully decided to split the content evenly: with an equal emphasis on both before and after the explosion and both theoretical and observational aspects. \\

In total, there were 15 invited review talks covering topics such as binary star evolution, stellar nucleosynthesis, the progenitors of supernovae, the possible explosion mechanisms and the nature of superluminous supernovae. These talks provided a survey of the current status of these fields and their connections:

Binary stellar evolution has been shown to be particularly important in the evolution of massive stars, with most stars that are expected to explode as supernovae having undergone some interaction with a binary companion ({\bf S. De Mink}).  This interaction with a binary companion can have major consequences on the mass loss history of the exploding star and the nature of the subsequent supernova, including the relationship with more exotic events such as superluminious supernovae and gamma ray bursts ({\bf Ph. Podsiadlowski}).  Modelling the evolution of massive stars is extremely complicated and key uncertainties remain, especially at the later stages when these stars are evolving more quickly and undergo brief periods fuelled by extreme nuclear reactions.  The latest generation of models need to consider 3D effects, derived from hydrodynamical models, such as rotation, dynamical instabilities and magnetic fields ({\bf R. Hirschi}).

The complex nature of stellar evolution modelling, and how different physical scenarios are handled, has major implications for the predictions of the final observed state of stars immediately prior to the supernova explosion; in particular the types of evolutionary phases different types of stars, with or without a binary companion, will go through and the type of supernova that will result ({\bf J. Groh}).  Theoretical models of massive star evolution are confronted by observations of massive stars; especially resolved stellar populations in the Milky Way and Local Group galaxies.  Stellar populations provide an overview of the various evolutionary stages that massive stars progress through,  as well as effects of metallicity, and the relationships between blue, yellow and red supergiants and Wolf-Rayet stars ({\bf P. Massey}).  The most common type of supernova in the local universe are the Type IIP events arising from red supergiants.  Intense examination of these stars, in particular, provides important clues to the key evolutionary stages massive stars undergo as they approach explosion ({\bf B. Davies}) and clues to the types of stars responsible for the supernovae we observe.

A key factor in the evolution of massive stars is the intense level of mass loss that they will experience as they evolve.  Strong stellar winds will dictate the mass of the star before it explodes, the type of the subsequent supernova and the formation of the circumstellar medium into which the supernova will eventually explode ({\bf J. Vink}).  Very massive stars ($M_{init} > 40M_{\odot}$) may undergo a phase of spectacular eruptive mass loss as Luminous Blue Variables for which, despite not being predicted to be direct progenitors of supernovae, there is growing evidence that such stars may be responsible for the interacting Type IIn supernova subtype.  The presence of dense shells of circumstellar material may result in some of the most luminous supernovae observed ({\bf N. Smith}).

A major advancement in our understanding of which stars will produce core-collapse supernovae has been the recent detections of the progenitor stars in fortuitous pre-explosion observations.  These have overwhelmingly shown red supergiants to be responsible for the Type IIP supernovae, but have had only a singular success in the identification of a progenitor of a hydrogen-poor supernova ({\bf S. Van Dyk}).

The explosion of a core-collapse supernova marks the birth of a neutron star or a black hole.  Models of these vital few seconds that signify the end of the life of the star require a mechanism to reverse the collapse of the star and, instead, result in its explosion.  Major models of these events rely on physical processes such as delayed neutrino heating and magnetohydrodynamic effects, but recent advancements all point towards these mechanisms all being fundamentally three-dimensional in nature, even with respect to the pre-collapse conditions ({\bf S. Couch}).  After the explosion, the resulting evolution of the supernova may provide key indicators to the nature of the progenitor through the evolution of the brightness with time and the presence of different chemical elements discerned through spectroscopy of the ejecta.  Such observations provide an alternative handle on the nature of the star prior to explosion, by probing its exploded remains ({\bf L. Dessart}).  

At late-times, during the optically thin nebular phase, it is possible to probe the entirety of the ejecta right down to the core. Such observations are able to place constraints on the 3D morphology of the ejecta and hydrodynamical instabilities arising from the explosion ({\bf C. Fransson}).  Despite the large distance towards most supernovae putting them beyond the spatial resolution capabilities of current and even planned telescope facilities, polarimetry is a unique observing technique which can observationally test the 3D predictions of the different explosion models ({\bf M. Tanaka}).  The ever increasing amount of observational data being acquired for supernovae and their host environments means that we are now entering an era where, whilst the physical properties for individual objects may not be clear, it is possible to build a broader statistical picture for the different supernova types ({\bf M. Modjaz}).

Recently, the observation of a new class of ``superluminous supernovae" has highlighted the role of new physical processes, such as the injection of energy from newly born magnetars, and the role of established processes, such as interaction with a dense circumstellar medium, in producing the large luminosities of these events ({\bf A. Gal-Yam}).\\

In addition, there were 8 contributed talks and 20 posters, reporting the latest results in the fields of massive star and supernova evolution, and discussion panels composed of early career researchers (kindly chaired by Lars Bildsten and Stuart Ryder).\\

The meeting was organised by Justyn R. Maund and H.-Thomas Janka (Supernovae) and Paul Crowther and Norbert Langer (Massive Stars).   The success of the meeting at Chicheley Hall continued and a year later, under the auspices of H.-Thomas Janka, many of the participants of ``Bridging the Gap" reconvened at Schlo{\ss} Ringberg in Tegernsee, Germany as part of the ``Progenitor-Supernova-Remnant Connection" workshop in July 2017.\\

\aucontribute{JRM wrote the original draft and THJ, PAC and NL contributed to subsequent iterations.}

\competing{The authors declare that they have no competing interests.}

\funding{The research of JRM is supported through a Royal Society University Research Fellowship.  THJ is supported by the Deutsche Forschungsgemeinschaft through the Excellence Cluster ``Universe" EXC 153 and by the European Research Council through grant
ERC-AdG No. 341157-COCO2CASA.}

\ack{The organisers would like to express their gratitude to the Royal Society for funding this meeting and hosting it in the magnificent surroundings of Chicheley Hall.  We are especially grateful to Naomi Asantewa-Sechereh, for her invaluable support in organising the logistics of this meeting, and to Bailey Fallon, for helping to arrange the publication of this volume.}



\end{document}